\documentclass[twocolumn,secnumarabic,amssymb,nobibnotes,aps,prl]{revtex4}
\usepackage{graphicx}
\usepackage{bm}
\newcommand{\bea}{\begin{eqnarray}}
\newcommand{\eea}{\end{eqnarray}}
\newcommand{\be}{\begin{equation}}
\newcommand{\ee}{\end{equation}}
\newcommand{\bt}{\begin{tabular}}
\newcommand{\et}{\end{tabular}}

\newcommand{\beas}{\begin{eqnarray*}}
\newcommand{\eeas}{\end{eqnarray*}}

\begin{document}
\date{\today}
\title{Phase Structure in a Hadronic Chiral Model}
\author{D.~Zschiesche$^a$, G.~Zeeb$^{a}$ and S.~Schramm$^{a,b}$}
\affiliation{$^a$Institut f\"ur Theoretische Physik, J.\ W.\ Goethe-Universit\"at,\\
Max-von-Laue-Str.\ 1, 60438 Frankfurt, Germany}
\affiliation {$^b$Center for Scientific Computing (CSC), J.\ W.\ Goethe-Universit\"at,\\
Max-von-Laue-Str.\ 1, 60438 Frankfurt, Germany}

\begin{abstract}
We study the phase diagram of a hadronic chiral flavor-SU(3)
model. Heavy baryon resonances can induce a phase structure that
matches current results from lattice-QCD calculations at finite
temperature and baryon density. Furthermore, we determine trajectories
of constant entropy per net baryon in the phase diagram.
\end{abstract}

\maketitle


Understanding hot and baryon-dense QCD matter is of central 
importance in theoretical and experimental heavy-ion physics.
Various effective theories for chiral
symmetry restoration predict that a line of first-order phase
transitions in the plane of quark-chemical potential $\mu_q$ versus
temperature $T$ ends in a critical point as the chemical potential is
lowered (see~\cite{Stephanov04} for a review).  Presently, the lattice
locates that point at $T\approx160$~MeV,
$\mu_q\approx120$~MeV~\cite{Fodor:2001pe}.  
The Compressed Baryonic
Matter (CBM) experiment at GSI FAIR is planned to 
perform a dedicated experimental effort to detect that line of
first-order phase transitions
in relativistic heavy-ion collisions. It is hoped that
by varying experimental parameters like the beam energy one could
trigger phase transitions of variable strength (latent heat) and
perhaps even locate the expected second-order critical point.

Models relying exclusively on order-parameter dynamics
typically predict significantly lower chiral phase transition
temperatures in baryon-dense matter than those found on the lattice
(see e.g.\ Fig.\ 6 in~\cite{Stephanov04}). As shown by Gerber and
Leutwyler some time ago~\cite{Gerber89}, while heavy hadronic states
are suppressed by the Boltzmann factor, their contribution to the
energy density at high temperature is substantial. This agrees with
other studies using a hadron resonance gas approach, which provides a
reasonable description of the thermodynamics obtained on the lattice
below the critical temperature \cite{bielefeld}.
Heavy states also reduce~\cite{Gerber89} the strong dependence of
the ``critical temperature'' (defined via the peak of a suitable
susceptibility) on the pion mass obtained in simple models for chiral
order parameter dynamics~\cite{Dumitru03}. The lattice indicates
a relatively weak dependence of $T_c$ on the pion mass~\cite{Karsch:2000kv}.

In this Letter we investigate the role of heavy
hadronic states on the location of the chiral critical point within a
non-linear $SU(3)_L\times SU(3)_R$ chiral model~\cite{model}. Here,
the phase transition at high temperature and baryon density is ''driven''
by baryonic resonance degrees of freedom, as suggested by the
discussion above. We shall show that the model is able to reproduce not
only the sketched qualitative phase structure but also the location of
the endpoint. The properties of the high mass states (masses and
couplings) are important for the actual location of the chiral phase
transition line~\cite{SZZ} in the plane of $T$ and $\mu_q$.

Lattice results show that the susceptibility peaks of 
the chiral condensate and of the Polyakov loop coincide 
at $\mu_q=0$ \cite{karsch02} which indicates
that for small $\mu_q$ those transition(s)
involve a coupling of the chiral dynamics to the gauge 
fields, see e.g.~\cite{polychiral}. 
However, within a matrix model for Polyakov loops, the effect of
$\mu_q>0$ on the critical temperature for
deconfinement is suppressed by $1/N_c$ \cite{DPZ}.
In contrast,
 $\mu$-effects on the chiral critical temperature could be
relatively strong. Indeed,
models which couple chiral fields to
Polyakov loops do suggest a decoupling of chiral and deconfinement dynamics
at large chemical potential, and that 
the largest increase in the Polyakov loop still appears at
$T \approx T_c (\mu=0)$, while the peak of the chiral susceptibility 
is shifted to significantly smaller $T$ \cite{polychiral}. 
Thus at large chemical potential a confined phase with partially
restored chiral symmetry might exist.


Our chiral hadronic $SU(3)$ Lagrangian incorporates the complete set of
baryons from the lowest flavor-$SU(3)$ octet, as well as
the entire multiplets of scalar, pseudoscalar, vector and axialvector
mesons \cite{model,chiralfint}.  In mean-field approximation, the
expectation values of the scalar fields relevant for symmetric nuclear
matter correspond to the non-strange and strange chiral quark
condensates, namely the $\sigma$ and its $s\bar{s}$ counterpart
$\zeta$, respectively, and further the $\omega$ and $\phi$ vector
meson fields. Another scalar isoscalar field, the dilaton $\chi$, is
introduced to model the QCD scale anomaly. However, if $\chi$
does not couple strongly to baryonic degrees of freedom then it remains
essentially ``frozen'' below the chiral transition.
Consequently, we focus here on the role of the
quark condensates.

Interactions between baryons and scalar (BM) or vector (BV)
mesons, respectively, are introduced as
\begin{eqnarray}
\label{L_BM+V}
{\cal L}_{\rm BM} &=&
-\sum_{i}   \overline{\psi}_i \left( g_{i\sigma}\sigma + g_{i\zeta}\zeta
\right) \psi_i
~,\\
{\cal L}_{\rm BV}
  &=& -\sum_{i}   \overline{\psi}_i
\left( g_{i\omega}\gamma_0\omega^0 + g_{i \phi}\gamma_0 \phi^0 \right) \psi_i
~,
\end{eqnarray}
Here, $i$ sums over the baryon octet ($N$, $\Lambda$, $\Sigma$, $\Xi$).
A term ${\cal L}_{\rm vec}$ with
mass terms and quartic self-interaction of the vector mesons is also added:
\begin{eqnarray}
{\cal L}_{\rm vec} &=& \frac{1}{2}
a_\omega \chi^2 \omega^2 + \frac{1}{2}
a_\phi \chi^2 \phi^2
+ g_4^{\,4} (\omega^4 + 2 \phi^4 )~. \nonumber
\end{eqnarray}
The scalar self-interactions are
\begin{eqnarray}
\label{L_0}
{\cal L}_0 &=& -\frac{1}{2} k_0 \chi^2 (\sigma^2+\zeta^2) + k_1
    (\sigma^2+\zeta^2)^2 + k_2 ( \frac{ \sigma^4}{2} + \zeta^4)
    \nonumber \\ & &{} + k_3 \chi \sigma^2 \zeta
 - k_4 \chi^4 - \frac{1}{4}\chi^4  \ln\frac{\chi^4}{\chi_0^{\,4}}
   +\frac{\delta}{3} \chi^4 \ln\frac{\sigma^2\zeta}{\sigma_0^{\,2} \zeta_0}
~.
\end{eqnarray}
Interactions between the scalar mesons induce the spontaneous
breaking of chiral symmetry (first line) and the scale breaking via
the dilaton field $\chi$ (last two terms).

Non-zero current quark masses break chiral symmetry explicitly in
QCD. In the effective Lagrangian this corresponds to terms such as
\begin{eqnarray}
{\cal L}_{\rm SB} &=& -\frac{\chi^2}{\chi_0^{\,2}}
\left[m_\pi^2 f_\pi \sigma + (\sqrt{2}m_K^2 f_K -
\frac{1}{\sqrt{2}}
m_{\pi}^2 f_{\pi})\zeta \right]~.
\end{eqnarray}

According to ${\cal L}_{\rm BM}$ (\ref{L_BM+V}), the effective
masses of the baryons,
$m_i^*(\sigma,\zeta)=g_{i\sigma}\,\sigma+g_{i\zeta}\,\zeta$\,,
are generated through their coupling to the chiral condensates,
which attain non-zero vacuum expectation values due to their
self-interactions \cite{model} in ${\cal L}_0$ (\ref{L_0}).
The effective masses of the mesons are obtained as the second derivatives of
the mesonic potential about its minimum.

The baryon-vector couplings $g_{i\omega}$ and $g_{i\phi}$
result from pure $f$-type
coupling as discussed in~\cite{model},
$g_{i\omega} = (n^i_q-n^i_{\bar{q}}) g_{8}^V$\,, 
$g_{i\phi}   = -(n^i_s-n^i_{\bar{s}}) \sqrt{2} g_{8}^V$\,,
where 
$g_8^V$ denotes the vector coupling of the baryon
octet and 
$n^i$ the number of constituent quarks of species $i$ in a given hadron. 
The resulting relative couplings agree with additive quark model
constraints.

All parameters of the model discussed so far are fixed by either
symmetry relations, hadronic vacuum observables or nuclear matter
saturation properties (for details see \cite{model}).  
In addition, the
model also provides a satisfactory description of realistic
(finite-size and isospin asymmetric) nuclei and of neutron stars
\cite{model,chiralsu3t0}.
%

If the baryonic degrees of freedom are restricted to the members
of the lowest lying octet,
the model exhibits a smooth decrease of the chiral condensates
(crossover)
for both high $T$ and high $\mu$  \cite{model,chiralfint}.
However, additional baryonic
degrees of freedom
may change this into a first-order phase transition
in certain regimes of the $T$-$\mu_q$ plane, depending on the
couplings \cite{Thei83,chiralfint,SZZ}.
To model the influence of such heavy baryonic states, we add a single resonance
with mass
$m_R = m_{0} + g_R \sigma$
and vector coupling
$g_{R\omega} = r_V g_{N\omega}$\,.
The mass parameters, $m_{0}$\,, $g_R$  and the relative
vector coupling $r_V$ represent free parameters, adjusted
to reproduce the
phase diagram discussed above\footnote{Note that instead of an explicit mass term we could have
coupled the resonance to the dilaton $\chi$\,.}. 
In principle, of course,
one should include the entire spectrum of resonances.
However, to keep the number of additional
couplings small, we effectively describe all higher
baryonic resonances by a single state with
adjustable couplings, mass and degeneracy.


In what follows, the meson fields are replaced by their (classical)
expectation values, which corresponds to neglecting quantum and
thermal fluctuations. Fermions, of course, have to be integrated out
to one-loop. The grand canonical potential can then be written as
\begin{eqnarray}
\label{thermpot}
   {\Omega}/{V}&=& -{\cal L}_{\rm vec} - {\cal L}_0 - {\cal L}_{\rm SB}
-{\cal V}_{\rm vac} \\
& &{} -T \sum_{i \in B} \frac{\gamma_i }{(2 \pi)^3} 
\int d^3k \left[\ln{\left(1 + e^{-\frac 1T[E^{\ast}_i(k)-\mu^{\ast}_i]}\right)} \right]
\nonumber \\
& &{}+T \sum_{l\in M} \frac{\gamma_l}{(2 \pi)^3} 
\int d^3k \left[\ln{\left(1 - e^{-\frac 1T[E^{\ast}_l(k)-\mu^{\ast}_l]}\right)
}\right], \nonumber
\end{eqnarray}
where $\gamma_B, \gamma_M$ denote the baryonic and mesonic
spin-isospin degeneracy factors, respectively, and $E^{\ast}_{B,M} (k)
= \sqrt{{k}^2+{m_{B,M}^*}^2}$ are the corresponding single
particle energies. In the second line we sum over the baryon octet
states plus the
additional heavy resonance with degeneracy $n$ (assumed to be 16).
The effective baryon-chemical potentials are
$\mu^{\ast}_i = \mu_i-g_{i \omega} \omega-g_{i \phi} \phi$, with
$\mu_i= (n^i_q - n^i_{\bar{q}}) \mu_q + (n^i_s - n^i_{\bar{s}})
\mu_s$.  The chemical potentials of the mesons are given by the sum of
the corresponding quark and anti-quark chemical potentials.  The
vacuum energy ${\cal V}_{\rm vac}$ (the potential at $\rho_B=T=0$) has
been subtracted.

By extremizing $\Omega/V$ one obtains self-consistent gap equations
for the mesons. Here, we consider non-strange matter, adjusting
$\mu_s$ for any given $T$ and $\mu_q$
so that the net number of strange quarks
is zero. The dominant ``condensates'' then are
the $\sigma$ and the $\omega$ fields.



The properties of the QCD transition at vanishing chemical potential
depend on the number of quark flavors and on their masses.  In
quenched lattice gauge theory, which corresponds to infinitely heavy
quarks, the critical temperature is $T_c \approx 260~$MeV
\cite{Laermann03}.  Smaller quark or pion masses,
respectively, reduce the critical temperature~\cite{Karsch:2000kv}.
Lattice calculations with 2 and 3 flavors of light quarks predict
critical temperatures of $T_c\approx 175$~MeV and 
$T_c \approx 155$~MeV, respectively.
However, we note that recently \cite{katz_qm05} values as high as
$190~$MeV were reported for the 2-flavor case with reasonably small
quark masses, and for reduced pion over-degeneracy on the lattice.
For two flavors, the lattice results support a continuous transition
in agreement with symmetry arguments~\cite{Laermann03}.  For three
massless flavors, a first-order phase transition is expected, which
is washed out at a pion mass of $m_\pi \approx 70-260$~MeV
\cite{Laermann03}.

\begin{figure}[h]
\vspace*{-0.8cm}
\begin{center}
\includegraphics[width=7cm]{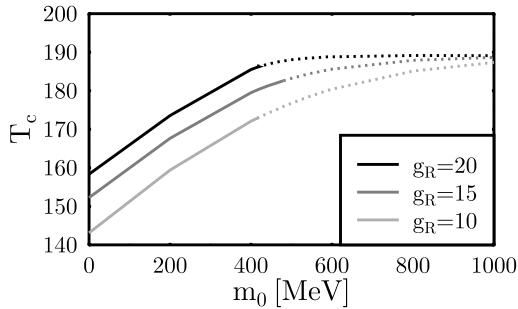}
\vspace{-0.5cm}
\caption{Critical temperature $T_c$ at vanishing chemical potential
vs the
explicit mass term $m_0$
for different values of the
scalar coupling $g_R$.
Full lines denote a first-order phase transition and
dotted lines correspond to a crossover.
\label{mu0}}
\end{center}
\vspace{-1cm}
\end{figure}
Within our model, at zero baryon density we obtain values for
the critical temperature around $140 - 190$ MeV, which are in
reasonable agreement with the numbers obtained from lattice
calculations.  Furthermore, we observe a rather slow increase of the
critical temperature with increasing explicit mass term $m_0$ 
(corresponding to the quark mass term of QCD, cf.\ Fig.\ \ref{mu0}). 
Finally, the first-order phase transition observed
for small explicit mass term 
indeed turns into a crossover
beyond some critical value of 
$m_0$. 
Neglecting the octet hyperons, which basically corresponds to the
two-flavor limit, leaves a crossover at $\mu=0$ and increases
the critical temperature.
If in turn we
assign the additional resonance to a spin-$3/2$ decuplet of flavor-$SU(3)$
and take the 3-flavor limit then chiral restoration happens via a
first-order phase transition and the critical temperature drops
by $\approx 30$~MeV. 
This behavior is also in (qualitative) 
agreement with the lattice QCD findings described above.

\begin{figure}[h]
  \vspace{-0.8cm}
   \begin{center}
    \includegraphics[width=08cm]{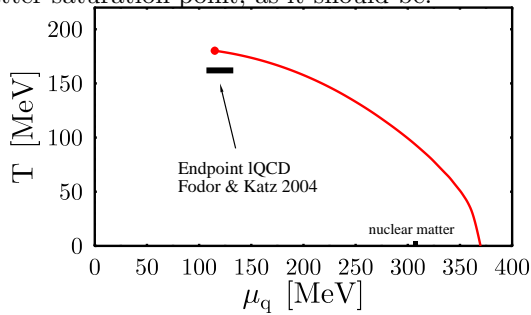}
  \vspace{-0.7cm}
    \caption{  \label{fig_pd_Testres}
 First-order phase transition line in the $T$-$\mu$ plane.
 $m_R(\sigma_0)=2$~GeV, $r_V=0.4$ and $m_{0} = 0.57$~GeV.  }
   \end{center}
  \vspace{-0.8cm}
  \small
 \end{figure}

We now turn to non-zero (net) baryon density.  Within certain ranges
for the mass and couplings of the heavy resonance, the location
of the critical endpoint is reasonably close to the lattice results. 
This is in contrast to
models which do not account for heavy degrees of freedom and usually
predict the endpoint at too low temperature (see
e.g.\ \cite{Stephanov04} for a summary).
Choosing, for example, $m_R(\sigma_0)=2$~GeV, $r_V=0.4$ and $m_{0} = 0.57$~GeV
results in $T_E\approx180$~MeV (see Fig.\ \ref{fig_pd_Testres}). This is a
little higher than $T_E$ from Ref.~\cite{Fodor:2001pe}
but the peak of the
susceptibility of the order parameter at
$\mu=0$ is at $T_c=185.6$~MeV, in good agreement with the
recent lattice results at zero density.
Smaller explicit mass term $\sim m_0$ for the additional heavy
resonance moves the critical endpoint
closer to the temperature axis, finally turning the crossover
at $\mu_q=0$ into 
a first-order transition (cf.\ Fig.\ \ref{mu0})
if $m_0$ drops below a critical value.
We also point out that the phase transition at $T=0$ occurs
at a density above the nuclear matter saturation point, as it should
be. 

\begin{figure}[h]
     \vspace{-0.6cm}
   \begin{center}
    \includegraphics[width=7.6cm]{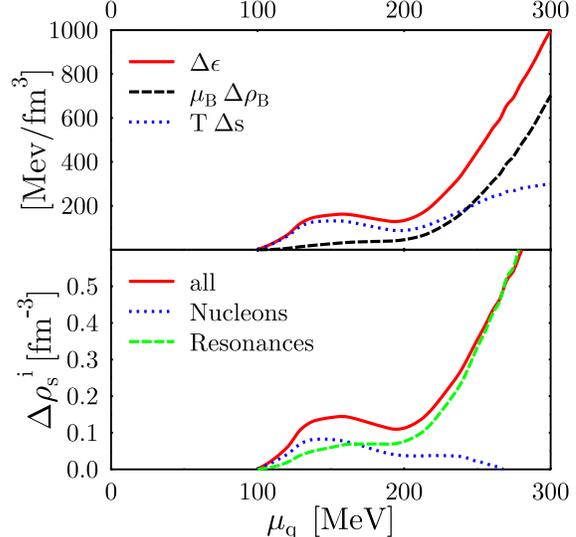}
     \vspace{-0.4cm}
    \caption{\label{latheat}
Top: Latent heat and its entropy ($T \Delta s$) and density
($\mu_B \Delta \rho_B$) contributions along the phase transition line
versus the quark chemical potential. Bottom: Difference of the scalar
density ($\Delta \rho_s^i$) in the two phases at the phase transition
versus the quark chemical potential. 
}
   \end{center}
  \vspace{-0.45cm}
  \small
 \end{figure}

Figure \ref{latheat} shows the density and entropy contributions
to the latent heat along the transition line of Fig.\ \ref{fig_pd_Testres}. 
The latent heat first grows with increasing chemical
potential. However, around $\mu_q = 
150$~MeV it decreases again, exhibiting a local minimum at $\mu_q
\approx 200$~MeV before increasing again towards low temperature. This
structure is also present in the entropy contribution.  The density
contribution to the latent heat up to $\mu_q
\approx 200$~MeV is flat but then rises very rapidly 
with increasing chemical
potential.

\begin{figure}[htb]
\centering
\vspace*{-.4cm}
\includegraphics*[width=80mm]{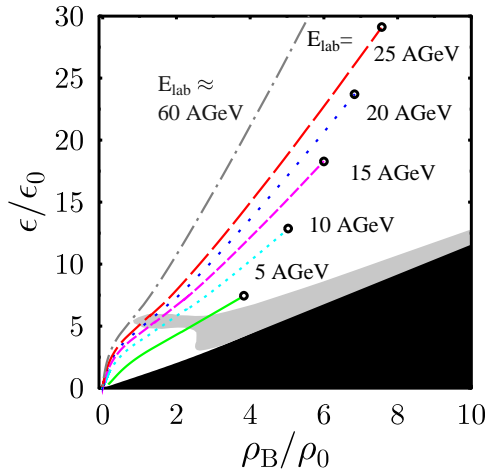}
\vspace*{-0.8cm}
\caption{Adiabats in the $e$-$\rho_B$ phase diagram. The grey area
represents the mixed phase.}
\label{erho}
\vspace*{-0.2cm}
\end{figure}

This somehow unexpected behaviour is due to a change in the 
dominant degrees of freedom. Around
$\mu_q\approx100-150$~MeV 
nucleons dominate the discontinuity of
the scalar density 
(cf.\ Fig.\ \ref{latheat}) at the phase boundary, but this contribution 
decreases above
$\mu_q \approx 150$~MeV and finally vanishes at $\mu_q\approx270$~MeV.
In contrast, the contribution of the resonances
increases sharply around 200 MeV. Both effects together generate 
the small dip in the latent heat and the 'nose' in the $e$-$\rho_B$
phase diagram, cf.\ Fig.\ \ref{erho}. 
This behaviour is actually not that surprising given that lattice results
at $\mu_q\approx0$ suggest just one single 
peak in all susceptibilities. In contrast, at
zero temperature the jump from zero to finite nucleon 
density signals the liquid-gas phase transition while the chiral
transition is expected to happen at larger chemical potential
\cite{Halasz}, driven
by newly populated heavier degrees of freedom. 


\begin{figure}[htb]
\centering
\vspace*{-.4cm}
\includegraphics*[width=7.5cm]{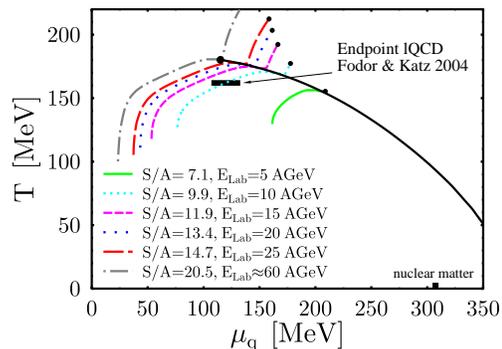}
\vspace*{-0.8cm}
\caption{Phase transition line and adiabats in the $T$-$\mu_q$ phase diagram.}
\label{tmu}
\vspace{-0.3cm}
\end{figure}
Figure \ref{tmu} depicts lines of constant entropy per net baryon 
(adiabats) in the phase
diagram. These correspond to perfect-fluid expansion trajectories.
At the phase transition they ``bend'' slightly towards smaller chemical
potential.  
For the present EoS, the
$S/A\approx20$ adiabat goes right through the endpoint of first-order
transitions, higher specific entropies then correspond to crossovers.
To make contact with experiments, e.g.\ at
the upcoming GSI-FAIR
facility, we use a simple estimate 
for the initial baryon 
number and energy density:
$\rho_B^{ini} = 2\,\gamma_{\rm CM}\,\rho_0$,
$e^{ini} = \sqrt{s}\, \rho_0\, \gamma_{\rm CM}$\,.
This assumes that the entire initial beam energy and baryon number
equilibrates in a Lorentz-contracted volume determined by the overlap
of projectile and target in the center-of-mass frame. 
In the energy range of interest here, this simple model
reproduces the specific entropy from more realistic three-fluid
models quite well~\cite{3fSA}. 
Within this simple model beam energies between
5-10$A$~GeV are sufficient to overshoot the shaded phase coexistence
region (Fig.~\ref{erho}).
In the energy range from 5-25$A$~GeV,
adiabatic expansion leads to a
first-order phase transition back to the symmetry broken state
with a latent heat of $\approx 120-160$~MeV/$\rm{fm}^3$.
Higher collision energies lead to weaker (more ''critical'') first
order transitions. The critical end point of
the line of first-order transitions is reached for a beam energy of
about 60 AGeV, with higher energies leading to crossover transitions.


In summary, we have explored the influence of heavy baryon resonances
on the phase diagram of a flavor-$SU(3)$ hadronic chiral Lagrangian.
We find that, with appropriate couplings, the model is able to
reproduce the structure of the QCD phase diagram at finite density in
that a line of first-order phase transitions at high density
terminates in a critical point at $T_E\approx180$~MeV, 
$\mu_E\approx110$~MeV. The interplay of nucleon and resonance
contributions to the density may lead to a non-monotonic latent heat
along the phase boundary.

We thank A.~Dumitru and H. St\"ocker for useful discussions.
This work was supported by DFG, GSI and BMBF and
used computational resources provided by the CSC at the 
University of Frankfurt, Germany.

\bibliographystyle{h-elsevier2.bst}
\end{document}